\documentclass[sw]{agu2001}
\usepackage{graphicx}
\authorrunninghead{WHEATLAND}

\titlerunninghead{A statistical flare forecast}

\authoraddr{M. S. Wheatland,
School of Physics, University of Sydney, NSW 2006, Australia.
(m.wheatland@physics.usyd.edu.au)}

\begin{document}

%
%
%
%
%

%
%

\title{A Statistical Solar Flare Forecast Method}
%

%
%


\author{M. S. Wheatland}
\affil{School of Physics, University of Sydney, NSW 2006, Australia}

%
%


\begin{abstract}
A Bayesian approach to solar flare prediction has been developed, which
uses only the event statistics of flares already observed. The method is
simple, objective, and makes few ad hoc assumptions. It is argued that 
this approach should be used to provide a baseline prediction 
for certain space weather purposes, upon which other methods, incorporating 
additional information, can improve. A practical implementation of the 
method for whole-Sun prediction of Geostationary Observational Environment 
Satellite (GOES) events is described in detail, and is demonstrated for 4
November 2003, the day of the largest recorded GOES flare.
A test of the method is described based on the historical record of 
GOES events (1975-2003), and a detailed comparison is made with 
US National Oceanic and Atmospheric Administration (NOAA) 
predictions for 1987-2003. Although the NOAA forecasts incorporate a 
variety of other information, the present method out-performs the NOAA 
method in predicting mean numbers of event days, for both M-X and 
X events. Skill scores and other measures show that the present 
method is slightly less accurate at predicting M-X events than the NOAA
method, but substantially more accurate at predicting X events, which are
important contributors to space weather.

\noindent
{\bf Keywords:} forecasting; space weather; flares

\noindent
{\bf Index Terms:} 7519 Solar Physics, Astrophysics, and Astronomy: Flares;
7554 Solar Physics, Astrophysics, and Astronomy: X rays, gamma rays, and
neutrinos 
\end{abstract}

%
%

%

\begin{article}

%
%

\section{Introduction}

Large solar flares are associated with a variety of space weather
effects. Those effects which occur promptly motivate flare prediction, 
or forecasting. For example, soft X-ray enhancements 
due to large flares cause increased ionization of the upper atmosphere, 
which can cause short-wave radio fadeouts, and Australia's Ionospheric 
Prediction Service (IPS) issues flare predictions on this basis.\footnote{see
http://www.ips.gov.au} Delayed space weather effects allow the possibility 
of physical modelling, given the knowledge that the flare has occurred. 
In this case prediction of the flare itself may be less important.

Solar flare prediction remains in its infancy, because of a lack of
detailed understanding of the physical processes underlying flares 
[e.g., Priest and Forbes, 2002]. Existing prediction methods are
probabilistic. One popular approach relies on the McIntosh
optical classification of sunspots, which divides sunspot groups into 
60 classes based on three parameters [McIntosh, 1990; Bornmann and Shaw,
1994]. The historical rate of flaring for a given classification provides 
an initial estimate for the expected flaring rate of flaring of an observed 
sunspot group. This approach is the basis for predictions published by the
Space Environment Center of the US National Oceanic and Atmospheric 
Administration 
(NOAA)\footnote{see http://www.sec.noaa.gov/ftpdir/latest/daypre.txt}
as well as NASA [Gallagher, Moon and Wang, 2002]\footnote{see 
http://beauty.nascom.nasa.gov/arm/latest/} and IPS.
NOAA/SEC uses an `expert system' developed by McIntosh [1990], and adopted
in 1987. The associated code begins with the McIntosh classification but 
also incorporates additional information, including dynamical properties of 
spot growth, rotation and shear, magnetic topology inferred from sunspot 
structure, magnetic classification, and previous (large) flare activity. 
The method involves more than 500 decision rules including `rules of thumb'
provided by human experts.

A variety of properties of active regions are known to correlate with
flare activity, and in principle could be incorporated into predictions. 
Attention has focused on photospheric magnetic field
measurements. For example, Sammis, Tang and Zirin [2000] confirmed that 
most large flares occur in large, magnetically complex regions. Studies of
vector magnetograms have suggested the length of the strongly-sheared 
strong field region along a neutral line as a predictor of flares and 
coronal mass ejections (CMEs) [e.g., Hagyard, 1990; Falconer, 2001].
Rapid emergence of new magnetic flux is also often identified as being
associated with flare occurrence [e.g., Schmieder et al., 1994].
Recently Leka and Barnes [2003] examined the relationship between 
moments of quantities constructed from vector magnetic field maps and 
flaring.

There are many problems with existing methods of flare prediction.
One problem with classification-based approaches is that they
tend to ignore the variability in flaring rate within a class. A related
difficulty is that choices for classes are ad hoc, and possibly subjective.
For example, the McIntosh classification is an arbitrary construction --- 
other choices of the three parameters could be made, and different observers
might disagree with a given classification. Similar criticisms apply
to the inclusion of additional information, e.g.\ properties of an 
active region, in existing methods of prediction. The choice of properties 
is essentially arbitrary, and it is unclear how the information can be 
included in an objective way.

The best indicator to future flaring activity is past flaring activity
[e.g., Neidig, Wiborg and Seagraves, 1990]. (In the prediction literature, 
the tendency of an active region which has produced large flares in the 
past to produce large flares in the future is termed persistence.)
Recently Wheatland [2004a] presented a Bayesian approach to solar flare 
prediction which uses the observed history of large and small flares
together with the known phenomenological rules of flare occurrence to make
a prediction for flaring. This `event statistics' approach has the 
advantage that it depends
only on past flaring activity, and so avoids the ad hoc choices implicit in
other prediction methods. The method has been developed into a practical
automated prediction scheme for whole-Sun prediction of soft X-ray flares 
based on NOAA solar event lists for
the Geostationary Observational Environmental Satellites (GOES). The 
method has been tested on the historical catalog of GOES events, and a 
brief account of this test was given in Wheatland [2004b]. 

Because of its simplicity and relative objectivity, the event statistics 
method is well suited to providing a baseline forecast for 
whole-Sun flaring for space weather purposes. 
Other methods of prediction, incorporating additional 
information, may then be applied to improve upon this baseline. In this 
paper the method and its implementation are described in detail. 
Section~2 reiterates the simple theory of the method [Wheatland, 2004a].
Section~3 describes the practical implementation for whole-Sun prediction
of GOES events, and section~4 gives a detailed account of the results 
of the test of this implementation on historical GOES data, including 
comparison with NOAA predictions for 1987-2003. Section~5 presents a brief
summary and discussion.
 
%
%


%
%

\section{Event statistics method} 

It is well known that the size distribution of flares 
(i.e.\ the distribution of some measure of flare size, such as peak flux 
in soft X-rays) obeys a power-law distribution [e.g., Crosby, Aschwanden 
and Dennis, 1993]. For a given choice $S$ of the measure of size the 
distribution may be written
\begin{equation}\label{eq:N(S)}
N(S)=\lambda_1(\gamma-1)S_1^{\gamma-1}S^{-\gamma},
\end{equation}
where $N(S)dS$ is the number of events per unit time with size in the 
range $S$ to $S+dS$, the quantity $\lambda_1$ is the total rate of events 
above the size $S_1$, and $\gamma$ is the power-law index. 
The value of the power-law index depends on the measure of size
$S$. For flare peak fluxes in X-ray bands, $\gamma$ is generally 
found to be in the range $1.7-1.9$ [e.g.\ Drake, 1971; Hudson, Peterson 
and Schwartz, 1969; Hudson, 1991; Lee, Petrosian and McTiernan, 1995; 
Shimizu, 1995; Feldman, Doschek and Klimchuk, 1997; 
Aschwanden, Dennis and Benz, 1998]. 

Typically we are interested in the rate of occurrence of
large events. If we denote the large size of interest $S_2$, then the 
corresponding rate may be denoted $\lambda_2$, and according to the 
distribution~(\ref{eq:N(S)}), this rate is given in terms of the
rate $\lambda_1$ by
\begin{equation}\label{eq:rate_large}
\lambda_2=\lambda_1
  \left( \frac{S_1}{S_2}\right)^{\gamma-1}.
\end{equation}
 
The power-law size distribution is one phenomenological rule of flare
occurrence. A second rule is that on short timescales, flares appear to
occur as a Poisson process in time. On longer timescales flare occurrence
may be described as a time-dependent Poisson process [e.g., Biesecker,
1994; Wheatland, 2001]. Assuming Poisson statistics, the probability of 
at least one large flare within a time $\Delta T$ is
\begin{equation}\label{eq:prob_large}
\epsilon = 1-\exp(-\lambda_2 \Delta T).
\end{equation}

Equations~(\ref{eq:rate_large}) and~(\ref{eq:prob_large}) provide a naive
prediction. If the power-law index $\gamma$ is estimated, and the current 
rate $\lambda_1$ of small events is estimated, then $\epsilon$ is the 
required probability. The Bayesian generalization involves calculating 
a posterior probability distribution $P(\epsilon)$ for the unknown parameter
$\epsilon$, given the available data and relevant background information
[e.g., Box and Tiao, 1992]. 
Specifically we assume that the data are a sequence of $M$ events with sizes
$s_1,s_2,...,s_M$ (where $s_i\geq S_1$ for each $i$), observed during an
observation interval from $t=0$ to $t=T$. The events are assumed to occur 
at times $0\leq t_1<t_2<...<t_M\leq T$. We also assume that an 
interval of time from $t=T-T^{\prime}$ to $t=T$ has been identified during
which the mean rate of events appears to be constant. A number $M^{\prime}$
of events is assumed to occur during this interval, which has 
duration $T^{\prime}$. 
A practical solution to identifying this interval is provided by the 
Bayesian blocks procedure [Scargle, 1998] which is a Bayesian 
changepoint algorithm designed to identify times
of rate variation. This procedure is discussed in more detail below.
As shown in Wheatland [2004a], the power-law index may be approximated from
the data by the maximum likelihood value $\gamma^{\ast}$:
\begin{equation}\label{eq:gam_ML}
\gamma^{\ast}=\frac{M}{\ln \pi}+1, \quad
{\rm where} \quad
\pi=\prod_{i=1}^M\frac{s_i}{S_1},
\end{equation}
and then the posterior distribution for $\epsilon$, based on
$\gamma^{\ast}$, $M^{\prime}$ and $T^{\prime}$ is
\begin{eqnarray}
P(\epsilon ) &=&
  C\left[-\ln (1-\epsilon ) \right]^{M^{\prime}}
  (1-\epsilon )^{\left( T^{\prime}/\Delta T\right)
  \left(S_2/S_1\right)^{\gamma^{\ast}-1}-1} \nonumber \\
  &\times &
  \Lambda \left[-\frac{\ln (1-\epsilon )}{\Delta T}
  \left(\frac{S_2}{S_1}\right)^{\gamma^{\ast}-1} \right],
\label{eq:P(eps)}
\end{eqnarray}
where $\Lambda (\lambda_1)$ is the prior distribution for $\lambda_1$,
and $C$ is the normalization constant, determined by the requirement 
$\int_{0}^{1}P(\epsilon)d\epsilon=1$. The prior distribution 
$\Lambda (\lambda_1)$ describes the values we would assign to $\lambda_1$ 
in the absence of any data. For a given prior distribution, the mean of
the posterior distribution provides an estimate for
$\epsilon$, and the width of the distribution provides an estimate
of the associated uncertainty.

The approximation involved in using Equation~(\ref{eq:gam_ML}) is valid
provided the power-law index $\gamma$ is narrowly defined by the data, 
which is true for large $M$. For smaller numbers of events it is necessary 
to simultaneously infer $\gamma$ and $\epsilon$, and the details are in
Wheatland [2004a]. For the applications in Sections~3 and~4, 
Equation~(\ref{eq:P(eps)}) is found to be sufficiently accurate.
 

%
%

\section{Application to GOES events}

\subsection{Implementation}

The method has been applied to event lists constructed from 
Geostationary Observational Environmental Satellite (GOES) 
observations at $1-8$\AA . The relevant measure of size, $S$ is
the peak X-ray flux in the $1-8$\AA ~channel, and it should be noted 
from the outset that this flux includes background. The GOES peak flux 
is routinely used to classify flares, with moderate-sized flares 
labelled M class (i.e.\ having a peak flux exceeding 
$S_2=S_{\rm M}=10^{-5}\,{\rm W}\,{\rm m}^{-2}$), and big flares
labelled X class (peak flux exceeding 
$S_2=S_{\rm X}=10^{-4}\,{\rm W}\,{\rm m}^{-2}$). 
Hence we are interested in predicting M class and 
X class events. Predictions are made for the occurrence of at least one 
event within $\Delta T=1\,{\rm day}$ of the prediction time. We take 
the time of events to correspond to the time of the peak flux recorded 
in the NOAA event lists.

It is necessary to choose a threshold size $S_1$ above which event sizes
are assumed to be power-law distributed. 
Figure~1 shows the peak flux distribution of 
all GOES events for 1975-2003 from the historical event lists available
from the NOAA National Geophysical Data Center.\footnote{see
ftp://ftp.ngdc.noaa.gov} The upper panel shows the data plotted as a
probability density function, or pdf (with bins of one tenth of a 
decade), and the lower panel shows the data plotted as a cumulative 
distribution (without binning). This figure shows that the data is 
power-law distributed for 
large peak fluxes. At low peak fluxes there is a departure from power-law
behavior, which is due to the problem of identifying small events
against the time varying soft X-ray background. A nominal threshold
$S_1=4\times 10^{-6}\,{\rm W}\,{\rm m}^{-2}$ for power-law behavior has
been chosen, and is indicated by the vertical solid line. Also shown by
a thick line is the power-law model distribution 
$P(S)=(\gamma^{\ast}-1)S_1^{\gamma^{\ast}-1}S^{-\gamma^{\ast}}$, with the
the maximum likelihood index
$\gamma^{\ast}$ given by Equation~(\ref{eq:gam_ML}). For these data we find 
$\gamma^{\ast}\approx 2.15\pm 0.01$. We note that this power law index is 
significantly larger than indices quoted in the literature for soft X-ray
peak fluxes [e.g.\ Aschwanden, Dennis and Benz, 1998], which are in the
range 1.7-1.9. The reason is that the data used here is not background
subtracted. For the prediction purposes it is preferable not to background
subtract, because we wish to make a prediction about the flux including
background. The method outlined here requires only that the quantity $S$ 
is power-law distributed above the chosen threshold, which is confirmed by
Figure~1. We will return to this point in Section~5. 

\begin{figure}
\noindent\includegraphics[width=20pc]{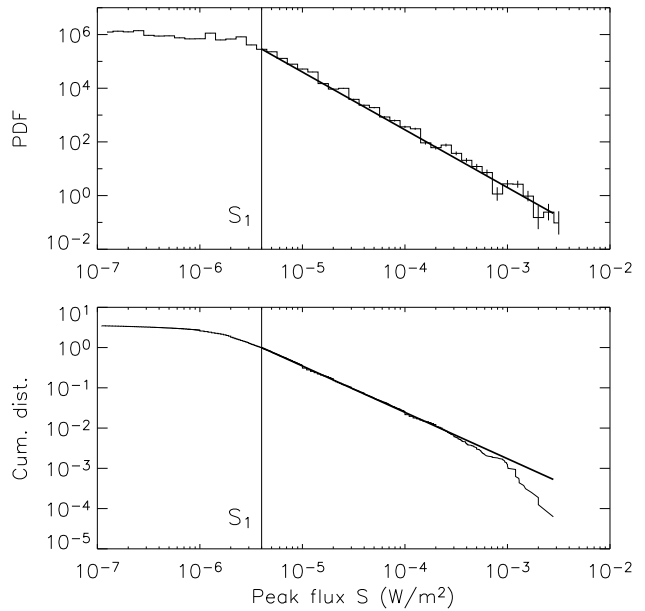}
 \caption{Upper panel: probability density function (pdf) for the
  peak flux ($1-8$\AA) of GOES events 1975-2003, together with 
  the threshold $S_1$ (vertical line) and the power-law model 
  (thick line). Lower panel: cumulative distribution.}
\end{figure}

Predictions are made using data within a window of duration $T$,
which is taken to span one year prior to the prediction time. 
Equation~(\ref{eq:gam_ML}) is applied to the year of 
events to determine $\gamma^{\ast}$. Then the Bayesian blocks procedure 
[Scargle, 1998] is applied to the year of data. This procedure takes the 
sequence of event
times $t_1,t_2,...,t_M$ and returns a sequence of changepoint times
$t_{B0}< t_{B1}<...<t_{BK}$ at which the rate is determined to change
(where $t_{B0}$ and $t_{BK}$ are the start- and end-time of
the data window), and a corresponding sequence 
$\lambda_{B1},\lambda_{B2},...,\lambda_{BK}$ of rates. 
The changepoints define $K$ `Bayesian blocks', i.e.\ intervals 
described by a single rate. The last Bayesian block provides $T^{\prime}$ 
and $M^{\prime}$ according to $T^{\prime}=t_{BK}-t_{B(K-1)}$ and 
$M^{\prime}=\lambda_{BK}T^{\prime}$.

The Bayesian blocks procedure involves Bayesian hypothesis testing, 
in which a single rate Poisson model is compared with a dual rate 
Poisson model for the data, for all possible choices of changepoints 
coincident with event times. If the dual rate model is more likely 
(by a factor \verb+PRIOR+, nominally taken to have the value two) 
the section of data is segmented, and the two segments are themselves 
subject to the test. This process is iterated. A segment is deemed 
complete when the single rate model is more likely, or when there is 
only one event in the segment. It should also be noted that the Bayesian
blocks procedure requires event times in discrete timesteps, and to this
purpose the peak times are rounded to the nearest minute.

The rates in the Bayesian blocks before the last block contain information
about how the flaring rate varies, and so are used to construct the 
prior $\Lambda (\lambda_1)$ in Equation~(\ref{eq:P(eps)}). The model
form
\begin{equation}\label{eq:prior_model}
\Lambda (\lambda_1)=a\exp (-b\lambda_1^c)
\end{equation}
was chosen for the prior, based on inspection of a rate distribution
constructed from a Bayesian blocks decomposition of the
GOES event lists for 1976-2003. For each prediction the parameters 
$a$, $b$, and $c$ are determined, for the given one-year window of 
data, by requiring that the first three moments of the model match the first
three moments of the data, estimated from the Bayesian blocks
results. Specifically we require
\begin{eqnarray}
\int_{0}^{\infty}d\lambda_1 \Lambda (\lambda_1)
  &=& 1, \nonumber \\
\int_{0}^{\infty}d\lambda_1 \lambda_1\Lambda (\lambda_1)
  &=& \sum_{i\neq K} N_{Bi}/\sum_{i\neq K}T_{B}i, \nonumber \\
\int_{0}^{\infty}d\lambda_1 \lambda_1^2\Lambda (\lambda_1)
  &=& \sum_{i\neq K}\lambda_{Bi}^2T_{Bi}/\sum_{i\neq K}T_{Bi},
\label{eq:moment_eqs}
\end{eqnarray}
where $N_{Bi}$ and $T_{Bi}=t_{Bi}-t_{B(i-1)}$ denote the number of
events in, and duration of, the $i^{\rm th}$ Bayesian block respectively,
where $\lambda_{Bi}=N_{Bi}/T_{Bi}$, and where the summations exclude 
the last block. As shown in the Appendix, these three conditions
uniquely determine values of $a$, $b$, and $c$.

Once $\gamma^{\ast}$, $M^{\prime}$ and $T^{\prime}$ have been determined,
and the prior has been constructed, Equation~(\ref{eq:P(eps)}) is used
(with $S_2=S_{\rm M}$ and $S_2=S_{\rm X}$) to construct posterior 
distributions $P_{\rm M}(\epsilon)$ and $P_{\rm X}(\epsilon)$ 
for the probability of occurrence of events above M size and above X size
respectively. The means of these distributions are taken as suitable 
estimates $\epsilon_{\rm M}$ and $\epsilon_{\rm X}$ of the probabilities 
of at least one M class event (or larger), and at least one X class event 
within $\Delta T=1\,{\rm day}$:
\begin{equation}
\epsilon_{\rm M} = \int_0^{1}d\epsilon \,\epsilon P_{\rm M}(\epsilon), 
\quad \quad
\epsilon_{\rm X} = \int_0^{1}d\epsilon \,\epsilon P_{\rm X}(\epsilon).
\label{eq:means}
\end{equation}
Corresponding uncertainties $\sigma_{\rm M}$ and $\sigma_{\rm X}$ 
may be obtained in the usual way from the first and second moments of the 
posterior distributions.
 
Another quantity of interest is the probability of getting at 
least one flare of M class, i.e.\ a flare with peak flux greater or 
equal to $10^{-5}\,{\rm W}\,{\rm m}^{-2}$, but less than 
$10^{-4}\,{\rm W}\,{\rm m}^{-2}$. To avoid ambiguity, we will refer to
events with peak flux in this range as M-X flares. 
If $\epsilon^{\prime}$ describes the 
probability of at least one M class event, or larger, and 
$\epsilon^{\prime\prime}$ describes the probability of at least one X class
event, or larger, then the desired probability is 
$\epsilon=\epsilon^{\prime}-\epsilon^{\prime\prime}$. 
If we denote the corresponding posterior distribution 
$P_{\rm M\!X}(\epsilon)$ then we have
\begin{eqnarray}
P_{\rm M\!X}(\epsilon)&=&\int_0^1d\epsilon^{\prime}\,
  \int_0^1d\epsilon^{\prime\prime}
  \delta(\epsilon-\epsilon^{\prime}+\epsilon^{\prime\prime})
  P_{\rm M}(\epsilon^{\prime})P_{\rm X}(\epsilon^{\prime\prime})  
  \nonumber \\
  &=& \int_0^1d\epsilon^{\prime} \,P_{\rm M}(\epsilon^{\prime})
  P_{\rm X}(\epsilon^{\prime}-\epsilon).
\label{eq:P(mu)}
\end{eqnarray}
The estimate for this quantity is taken to be the mean of the posterior
distribution, which is denoted $\epsilon_{\rm M\!X}$, and similarly the
associated uncertainty is denoted $\sigma_{\rm M\!X}$. 

\subsection{Example: application to 4 November 2003}

To illustrate the application of the method, we consider a prediction
for the time 00:00UT on 4 November 2003. The largest soft X-ray flare of the 
period of GOES observations (1975-2004) occurred starting at 19:29UT 
on that day. The event saturated the GOES detectors, but was estimated 
by NOAA to be X28, i.e.\ to have a peak flux of 
$2.8\times 10^{-3}\,{\rm W}\,{\rm m}^{-2}$ in the 1-8\AA ~band. The 
enhanced X-ray flux due to this flare caused a substantial lowering of the  
D-region of the ionosphere, suggesting an even higher peak flux of 
around X45 [Thomson, Rodger and Dowden, 2004].

Figures~2, 3 and~4 illustrate the prediction process for this day.

Figure~2 shows the size distribution of the one-year window of data
(prior to 4 November 2003 00:00UT) used in the prediction. Only events 
above the threshold flux $S_1=4\times 10^{-6}\,{\rm W}\,{\rm m}^{-2}$ 
(which is indicated by a vertical line) are shown, and there are 480 
events in all. The upper panel shows the pdf, and 
the lower panel the cumulative distribution. The power-law model 
(with maximum likelihood index $\gamma^{\ast}\approx 2.07\pm 0.05$) is 
indicated by a thick line. Once again we note that this value is
significantly larger than other values for the distribution of soft 
X-ray peak flux quoted in the literature, because of the lack of 
subtraction of background flux.

\begin{figure}
\noindent\includegraphics[width=20pc]{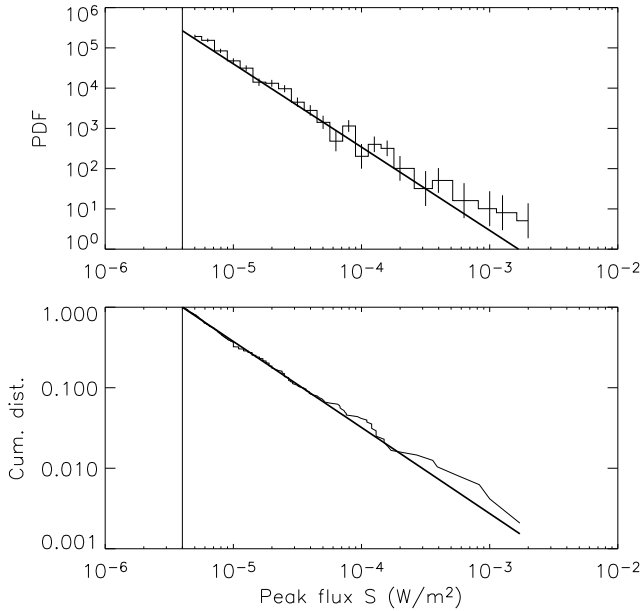}
 \caption{Upper panel: pdf for peak flux of $1-8$\AA ~GOES 
  events above threshold, for one year prior to 4 November 2003. 
  Lower panel: cumulative distribution.}
\end{figure}

Figure~3 illustrates the inference on the rate. The upper panel shows the
480 events (indicated by crosses) above the threshold during the one year 
window, as a plot of peak flux versus time. 
The lower panel shows the result of applying the
Bayesian blocks procedure to these data. The procedure decomposed the data
into 13 blocks. The last block had a duration 
$T^{\prime}\approx 15.3\,{\rm days}$ and contained $M^{\prime}=104$ 
events.

\begin{figure}
\noindent\includegraphics[width=20pc]{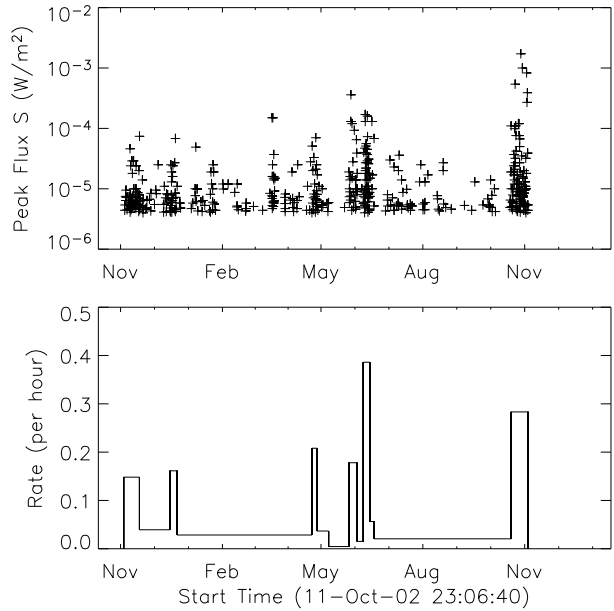}
 \caption{Upper panel: Peak flux of $1-8$\AA ~GOES events (crosses) 
  above threshold versus time, for one year prior to 4 November 2003. 
  Lower panel: Bayesian blocks decomposition of rate versus time.}
\end{figure}

Figure~4 shows the posterior distributions $P_{\rm M\!X}(\epsilon)$ 
(upper panel) and $P_{\rm X}(\epsilon)$ (lower panel). The estimates
$\epsilon_{\rm M\!X}$ and $\epsilon_{\rm X}$ obtained from the means 
of the distributions are shown as vertical lines. The result is that the
probability of at least one M-X event within one day is 
$\epsilon_{\rm M\!X}\approx 0.73\pm 0.03$, and 
the probability of at least one event of X size within one day is 
$\epsilon_{\rm X}\approx 0.19\pm 0.02$. These values are quite high, 
reflecting the high rate of flaring immediately prior to 4 November. 
However, the
results also show the limitations of probabilistic forecasting --- the 
prediction for an X class event is only around 20\%, yet the largest 
flare of the last three decades is imminent. 

\begin{figure}
\noindent\includegraphics[width=20pc]{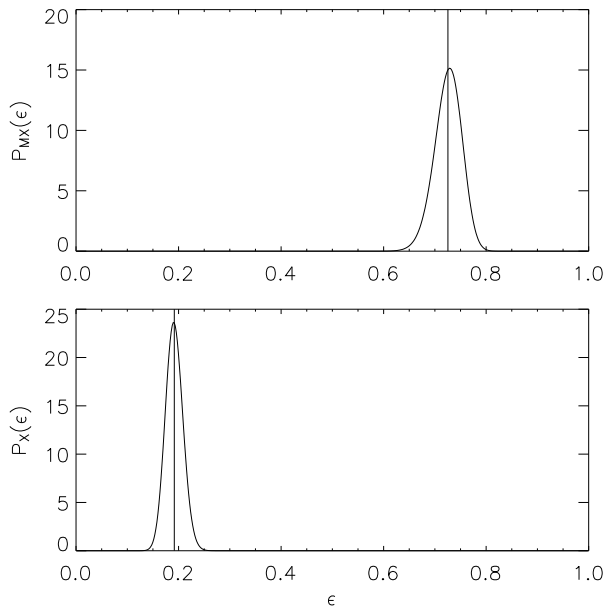}
 \caption{Upper panel: Posterior distribution for probability of 
  M-X events for 4 November 2003.
  Lower panel: Posterior distribution for X events for that day.}
\end{figure}

\section{Test of the method}

\subsection{Results}

To test the method, predictions were performed for each day of GOES
event data for 1975-2003. Each prediction was compared with the
historical fact of whether flares did or did not occur on the given 
day. A brief description of the test was given in Wheatland (2004b). Here
the test is described in more detail, including comparison, in Section~4.2, 
of the prediction statistics with published NOAA predictions for 
1987-2002.

Predictions were made according to the procedure outlined in 
Section~3.1. Since the predictions use a one-year window of data, 
the results are only considered for 1976-2003, i.e.\ the predictions 
for 1975 are omitted from consideration because they are based on 
less than a year of previous data.

Figure~5 summarizes the results, as plots of yearly numbers of observed
event days, i.e.\ days with one or more events (solid histograms), and 
yearly numbers of predicted event days (diamonds). The upper panel shows
the results for M-X events, and the lower panel shows the results for X
events. The numbers of predicted event days are the sums of the 
prediction estimates $\epsilon_{\rm M\!X}$ and $\epsilon_{\rm X}$ over 
all days in a given year. Representative error bars are shown for the
observations, corresponding to the square root of the number of events.
These error bars reflect the expected variability in the observed numbers.
These plots suggest that the method does quite well in predicting overall
event numbers, although some systematic over-prediction is apparent.

\begin{figure}
\noindent\includegraphics[width=20pc]{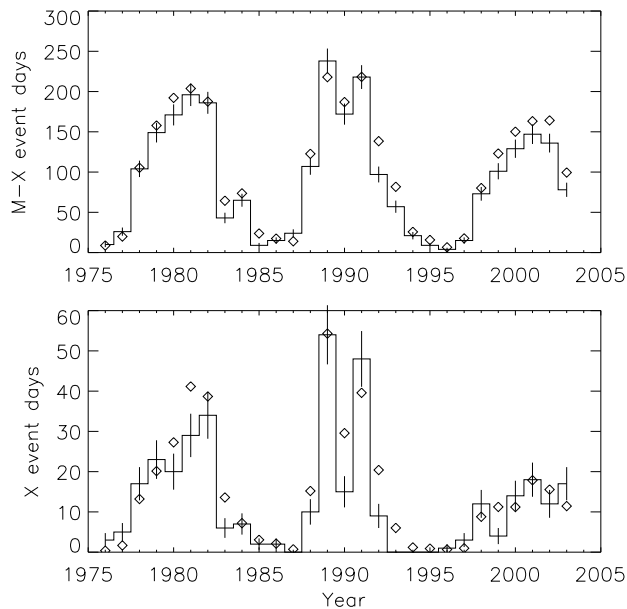}
 \caption{Comparison of predictions and observations for 1976-2003.
  Upper panel: Events days for M-X events:
  observed (histogram); predicted (diamonds). Lower panel: the same,
  but for X events.}
\end{figure}

Table~1 summarizes the statistics of the predictions for 1976-2003, 
in a notation following meteorological practice [e.g., Murphy and Winkler, 
1987]. To assess probabilistic predictions, the joint probability
distribution for forecasts (denoted $f$) and observations (denoted $x$)
may be constructed, and properties of this distribution examined. In the
present context we have $f=\epsilon_{\rm M\!X}$ or $f=\epsilon_{\rm X}$, 
for predictions for M-X or X events respectively. 
The value of $x$ for each day is zero or one, according to whether an event 
did or did not occur. Averages over all days are denoted by 
$\langle \cdots \rangle$. For example, $\langle f\rangle$ is the average
of the forecast probability over all days. As a second example, 
$\langle f | x=1\rangle$ is the average of the forecast probability over
all days on which at least one flare did occur. 
Standard deviations over all days are denoted by $\sigma$. 
`MAE' denotes the mean absolute error:
\begin{equation}
{\rm MAE}(f,x)=\langle |f-x|\rangle, 
\end{equation}
and `MSE' denotes the mean square error:
\begin{equation}
{\rm MSE}(f,x)=\langle (f-x)^2\rangle.
\end{equation}
The linear association is the correlation of $f$ and $x$. 

Table~1 also
gives the climatological skill score [e.g., Murphy and Epstein, 1989],
defined by
\begin{eqnarray}
{\rm SS}(f,x)&=&1-{\rm MSE}(f,x)/{\rm MSE}(\langle x\rangle,x) \nonumber \\
&=& 1-{\rm MSE}(f,x)/\sigma_x^2,
\label{eq:skill}
\end{eqnarray}
which is a measure of the improvement of the forecasts over a constant
forecast given by the average. Perfect prediction ($f=x$) corresponds to a
skill score of unity. A positive skill score indicates better performance,
and a negative skill score worse performance, with respect to the average.

The table indicates that the method performs quite well in describing
the overall frequency of occurrence of flares: we have 
$\langle \epsilon_{\rm M\!X}\rangle \approx 0.282$, whereas 
M-X events occurred on a fraction 0.254 of days; and  
$\langle \epsilon_{\rm X}\rangle \approx 0.040$, whereas X class events 
occurred on a fraction 0.036 of days. However, these values also 
confirm that there is a tendency for over-prediction.

The method also has good discrimination, i.e.\ it assigns substantially
higher values to $f$ on days on which events occurred, compared with
non-event days. The skill scores for the method are 0.272 (for M-X events) 
and 0.066 (for X events).

Another way to summarize the results is in terms of `reliability plots,' 
which show the success of the predictions as a function of 
$\epsilon_{\rm M\!X}$ or $\epsilon_{\rm X}$. Figure~6 shows the 
reliability plot for M-X events. This diagram is constructed as follows. 
The predictions 
$\epsilon_{\rm M\!X}$ for all days are sorted into bins of width 0.05. 
For each bin, the observed number of those days on which at least one 
event did occur is used to estimate the underlying probability of an event
on those days, and this is the vertical value for the bin.
Specifically, the estimate used is the Bayesian estimate assuming 
binomial statistics and a uniform prior: if there 
are $R$ days with at least one event out of a total of $S$ days, then 
the estimate for the probability is $p=(R+1)/(S+2)$ (Laplace's rule 
of succession), and the associated uncertainty is 
$[p(1-p)/(S+3)]^{1/2}$ [e.g., Jaynes, 2003]. 
On a reliability plot, perfect prediction corresponds to a 45 degree 
line, which is indicated by the solid line in Figure~6. This figure 
shows that the method has performed quite well for 
prediction of M-X events. There is some over-prediction (the points fall
below the perfect prediction line) for days on which the forecast has 
moderate values ($\epsilon_{\rm M\!X}\approx 0.25-0.65$).

\begin{figure}
\noindent\includegraphics[width=20pc]{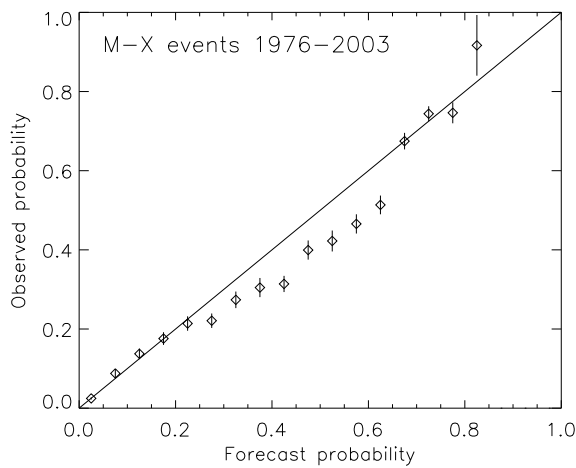}
\caption{Reliability plot for prediction of M-X events for 1976-2003. 
  The horizontal axis shows the probabilities assigned in the predictions, 
  and the vertical axis shows probabilities derived from the observed 
  frequencies of event occurrence.
 }
\end{figure}

Figure~7 shows the reliability plot for prediction of X events for
1976-2003. The predictions are conservative, in that 
$\epsilon_{\rm X}$ is less than about 0.5 for all days. Once again the 
method appears to perform quite well, with some tendency to 
over-prediction.

\begin{figure}
\noindent\includegraphics[width=20pc]{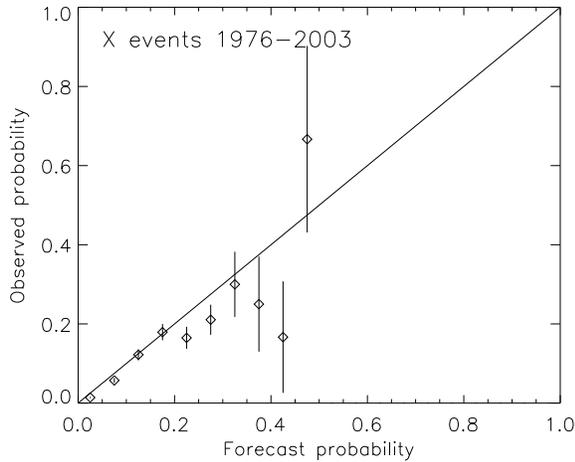}
 \caption{Reliability plot for prediction of X events for 1976-2003 (see
  Figure~6).}
\end{figure}

Although we have argued that the present prediction method involves
relatively few ad hoc choices, there are a number of free parameters,
in particular \verb+PRIOR+, the prior ratio involved in the Bayesian 
blocks routine, $S_1$, the threshold for the power-law model, 
$T$, the length of the data window used in the predictions, as well as 
the functional form~(\ref{eq:prior_model}) for the prior rate 
distribution. It is interesting to briefly examine the effect on the 
predictions of varying some of these choices. Table~2 shows the effect on
$\langle \epsilon_{\rm M\!X}\rangle$ and $\langle \epsilon_{\rm X}\rangle$ 
of varying these choices one by one. The observed means are given at the 
bottom. This table suggests that the method is relatively insensitive to the 
choices \verb+PRIOR+ and $T$. If the threshold $S_1$ is halved, the 
predictions are higher, and hence worse. This effect is probably due to 
the departure from power-law behavior at small sizes (see Figure~1) 
causing the inferred power-law index to be too small. Table~1 also 
indicates that if the prior for the rate is ignored [corresponding to 
the choices $a=1$, $b=0$, $c=1$ in Equation~(\ref{eq:prior_model})] 
the predictions are worse. This shows that the Bayesian blocks before the
last are providing useful information for prediction.

\subsection{Comparison with NOAA predictions}

As described in Section~1, the NOAA uses the McIntosh expert system
[McIntosh, 1990] to make flare predictions.
The NOAA publishes web pages with tables of statistics describing the 
reliability of its flare forecasts for the period 
1987-2003.\footnote{see http://www.sec.noaa.gov/forecast\_verification/}
Using these tables, it is possible to compare the NOAA predictions with 
those of the present method. 

Table~3 compares the predictions of the two methods for 1987-2003. 
The table lists the means of the forecast probabilities
$f=\epsilon_{\rm M\!X}$ and $f=\epsilon_{\rm X}$ for the present method 
and for the NOAA method, as well as the means of the observed values $x$. 
The present method gives mean prediction probabilities closer to the 
observations for both M-X and X events. 
For M-X events, both methods show similar over-prediction, although the 
present method is slightly better. For X events the present method gives 
substantially improved mean prediction probabilities compared with
the NOAA method. 

The average of $f$ is an incomplete measure of the success of a 
prediction method, since it ignores e.g.\ whether high predictions are 
assigned on event days, and low predictions on non-event days. 
Hence Table~3 also lists the average forecasts for event days and non-event
days, the mean absolute error, the mean square error, and the skill scores,
for the two methods. 
The values of $\langle f|x=1\rangle$ and $\langle f|x=0\rangle$ 
show that the NOAA method is somewhat more discriminating than the present
method, for both M-X and X event prediction. However,
the present method has a lower mean prediction for non-event days for X
class flares. The mean absolute and mean square errors suggest that the
NOAA method is slightly more accurate for M-X event prediction, but less
accurate for X event prediction. The overall skill scores also support
this. Notably the NOAA skill score for X event prediction is negative.

It is also interesting to compare the predictions and observations on a
year by year basis. Figure~8 shows the predicted numbers of event days 
for the present method (diamonds), the predicted numbers of event days for 
the NOAA method (asterisks), as well as the observed number of event days 
(solid histograms), for each year in the period 1987-2003. The upper panel 
shows the results for M-X events, and the lower panel for X events. The
predictions for M-X events for the two methods show a comparable scatter
around the observed values. The predictions for X events are much better in
the case of the present method, in particular for cycle 23.

\begin{figure}
\noindent\includegraphics[width=20pc]{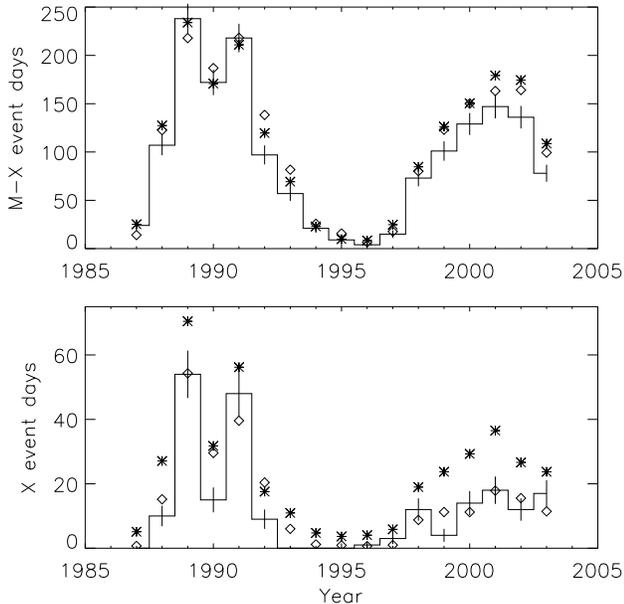}
 \caption{Comparison of observed event days (solid histograms), the 
predictions of the present method (diamonds), and NOAA predictions
(asterisks), for 1987-2003. Upper panel: M-X events. Lower panel: 
X events.}
\end{figure}

Figure~9 compares the skill scores [Equation~(\ref{eq:skill})] for the
predictions by the two methods, on a year by year basis. The scores for the
present method are shown by diamonds, and the scores for the NOAA method by
asterisks, with the upper panel showing the results for M-X events, and 
the lower panel X events. The NOAA method is seen by this measure to be
slightly better at predicting M-X events but worse at predicting X events 
(and in particular had two years with very poor X event predictions).

\begin{figure}
\noindent\includegraphics[width=20pc]{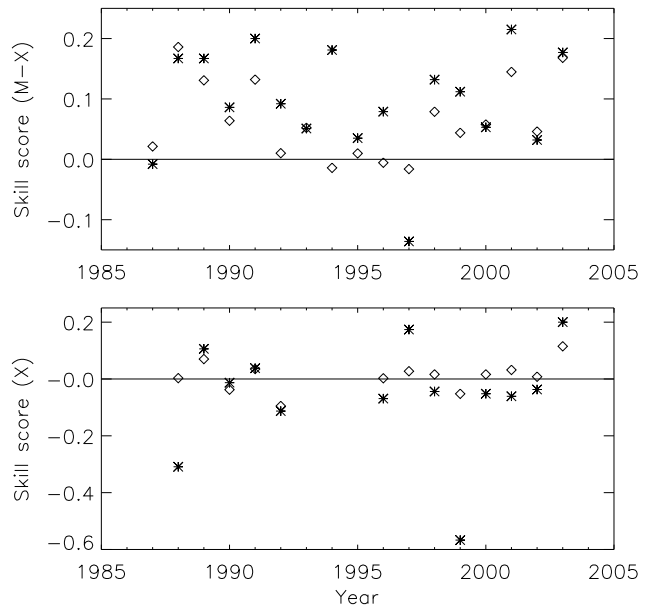}
 \caption{Comparison of skill scores for the present method (diamonds),
and for NOAA predictions (asterisks), for 1987-2003. Upper panel: M-X
events. Lower panel:X events.}
\end{figure}
\section{Discussion}

A practical implementation of an event statistics approach to solar flare
prediction [Wheatland 2004a] is demonstrated for daily whole-Sun 
prediction of GOES soft X-ray flares, and is illustrated by application 
to 4 November 2003, the day of the largest recorded GOES flare 
(Figures~2, 3 and~4). The method makes predictions based only on the 
observed history of flaring by exploiting the phenomenological rules of 
flare occurrence, in particular the power-law distribution of flares in 
size (Figure~1). The method is 
simple, both conceptually and in terms of implementation, and involves few
ad hoc assumptions. It requires no specialized observations, only 
the NOAA solar event lists. As such it is well-suited to 
providing a first guess for the probability of flare occurrence.
In principle additional observations of
physical parameters related to flaring could be used to improve upon
the basic prediction provided by the present method, and approaches 
to this problem will be considered in future work.

The method has been tested on prediction of GOES events for each day in the
period 1976-2003. The test was briefly described in Wheatland [2004b],
and a detailed account is given here. 
The method is found to perform well in assigning probabilities 
to the occurrence of events in the range M to X (`M-X events') and to 
the occurrence of X events, although there is a tendency to 
over-prediction, in particular for M-X events. 
A number of measures of success of the method are examined, including
predicted numbers of event days (Figure~5) reliability plots 
(Figures~6 and~7), and verification statistics (Table~1). 
In particular, the skill scores for M-X event prediction and X event 
prediction are found to be 0.272 and 0.066 respectively. 

The results of the test are found to only weakly depend on a number of 
chosen parameters in the method, in particular a prior ratio for segmentation
in the the Bayesian blocks procedure used to determine the current rate,
and the length of the data window used (see Table~2). 
However, the predictions require an accurate choice of the threshold for 
the power-law size distribution, and benefit from the inclusion of prior 
information on the rate.
  
The present method is also compared with the long-standing NOAA prediction 
method, using the NOAA's published prediction statistics for 
1987-2003 (Table~3 and Figures~8 and~9). The event statistics method is found 
to out-perform the NOAA method in predicting overall numbers of M-X and 
X event days. In particular the NOAA method is observed to seriously 
over-predict X class events (lower panel, Figure~8). The NOAA method is
found to be slightly more accurate in M-X event prediction, but less
accurate in X event prediction, based on skill scores and other validation
statistics (Table~3). Overall the present method provides improved
prediction of X class flares [e.g.\ compare the skill scores 
${\rm SS}(f,x)$ in Table~3]. This is significant because X flares,
although infrequent, are the most important flares from the point of 
view of space weather. It is perhaps surprising that the present method
fares as well as it does, given the amount of background information 
incorporated into the NOAA forecasts [McIntosh, 1990]. 
These results support the contention that the event statistics method 
is well suited to providing a baseline forecast, upon which other methods 
can improve. 

The tendency to over-predict moderate sized flares is still being 
investigated, but it is likely that it stems from the 
determination of the current rate. The Bayesian blocks procedure is
always trying to `catch up' with variations in the Sun's flaring 
rate. If the Bayesian blocks method is
systematically late in detecting a sudden decline in rate (e.g.\ due to the
decay of an active region, or its rotation off the disk), then there will
be a period of over-prediction. A specific issue with the implementation
of the Bayesian blocks procedure is that there must be at least one event
in a block, so that the inferred rate is never identically zero, even if
the true rate is zero. This is expected to lead to overestimation of the
rate at times of low activity. A related problem is that it is intrinsically 
difficult to accurately determine low rates because of the absence of events. 
In future work these questions will be examined
more carefully. It should also be noted that the Bayesian blocks procedure
is not guaranteed to find the optimal decomposition [Scargle, 1998].
Recently Scargle described a new, optimal Bayesian blocks algorithm
[Scargle 2004], and in future the new method will be applied to flare
prediction.

In Section~3 it was noted that the $1-8$\AA ~GOES peak fluxes used 
here are not background subtracted. For many applications of soft X-ray
data, e.g.\ determining intrinsic properties of flares, it is essential 
to perform accurate background subtraction [e.g.\ Bornmann, 1990]. It 
is also important in statistical studies where the concern is with the 
distribution of the intrinsic quantity, for example in studies 
investigating what the distributions of peak flux reveal about underlying 
physical processes. However, in the present context background subtraction 
is unnecessary. As noted in Section~3, provided the peak fluxes (including
background) are power-law distributed, they are suitable for predictive
purposes. This is an advantage of the method, in that readily available
data (GOES event lists) may be used as the basis for a prediction. 
However, the lack of background subtraction means that the peak flux 
distributions constructed here cannot be readily compared with other 
published distributions. In particular, the power-law indices obtained 
here are typically larger than two, whereas a large body of literature 
reports that the intrinsic distribution of X-ray peak flux has an index 
in the range $1.7-1.9$ [e.g.\ Drake, 1971; Hudson, Peterson
and Schwartz, 1969; Hudson, 1991; Lee, Petrosian and McTiernan, 1995;
Shimizu, 1995; Feldman, Doschek and Klimchuk, 1997;
Aschwanden, Dennis and Benz, 1998].

Although convenient, the GOES event lists also have a specific shortcoming
for predictive purposes. Events are selected against a substantial and 
time-varying soft X-ray background. At times of high solar activity, more 
events are missed because of the increased background [e.g., Wheatland, 
2001], and this leads to a relatively large threshold for power-law 
behavior of the peak fluxes (see Figure~1), which is a disadvantage for 
the present method. In future other datasets will be considered. However, 
the advantages of the GOES event lists are their availability, and the 
close relationship of soft X-ray peak flux to an important space weather 
effect (ionization of the upper atmosphere).

Predictions made using the method described in this paper are now 
published daily on the web.\footnote{see 
http://www.physics.usyd.edu.au/$\sim$wheat/prediction/}
The web pages also include running measures of how accurate the 
published predictions are, in the form of automatically updated plots of 
reliability and skill scores. 
The codes used to make the predictions are written in the Interactive
Data Language (IDL)-based SolarSoft system [Freeland and Handy, 1998], 
which has become the de facto standard for solar data analysis.
All codes are available on request from the author.

\section*{Appendix}

The moments of the model distribution~(\ref{eq:prior_model}) are given
by
\begin{eqnarray}
\langle \lambda_1^{\alpha} \rangle & \equiv &
  a\int_{0}^{\infty}\lambda_1^{\alpha}\exp (-b\lambda_1^c )\,d\lambda_1
  \nonumber \\
  &=&\frac{a}{b^{(\alpha+1)/c}c}\Gamma \left(
  \frac{\alpha+1}{c}\right),
\label{eq:gamma_integral}
\end{eqnarray}
where $\Gamma (x)$ is the Gamma function. Hence the three 
moment equations~(\ref{eq:moment_eqs}) may be written
\begin{eqnarray}
\frac{a}{b^{1/c}c}\Gamma (1/c)&=&1 \nonumber \\
\frac{a}{b^{2/c}c}\Gamma (2/c)&=& \overline{\lambda_{1}} \nonumber \\
\frac{a}{b^{3/c}c}\Gamma (3/c)&=& \overline{\lambda_{1}^2},
\label{eq:moment_eqs2}
\end{eqnarray}
where $\overline{\lambda_1}$ and $\overline{\lambda_1^2}$ denote the
right hand sides of the second and third of 
equations~(\ref{eq:moment_eqs}).

Eliminating $a$ between the first and second of 
Equations~(\ref{eq:moment_eqs2}) gives
\begin{equation}\label{eq:moment_eqs3}
b^{-1/c}\frac{\Gamma (2/c)}{\Gamma (1/c)}=\overline{\lambda_1},
\end{equation}
and eliminating $a$ between the first and third of
Equations~(\ref{eq:moment_eqs2}) gives
\begin{equation}\label{eq:moment_eqs4}
b^{-2/c}\frac{\Gamma (3/c)}{\Gamma (1/c)}=\overline{\lambda_1^2}.
\end{equation}
Eliminating $b$ between Equations~(\ref{eq:moment_eqs3})
and~(\ref{eq:moment_eqs4}) gives the transcendental equation for $c$:
\begin{equation}\label{eq:moment_eqs5}
\left[ \Gamma (2/c)\right]^2\overline{\lambda_1^2}
  -\left(\overline{\lambda_1}\right)^2\Gamma (1/c)\Gamma (2/c)=0.
\end{equation}
This equation needs to be solved for $c$, e.g.\ using Newton-Raphson,
for a suitable initial guess, and then Equation~(\ref{eq:moment_eqs3}) 
gives $b$:
\begin{equation}
b=\left[\Gamma (1/c) \overline{\lambda_1}/\Gamma (2/c)\right]^{-c}
\end{equation}
and the first of Equations~(\ref{eq:moment_eqs2}) gives $a$:
\begin{equation}
a=b^{1/c}c/\Gamma (1/c).
\end{equation}

%
%


%

%


%
%

\begin{acknowledgments}
The author acknowledges the support of an Australian Research Council
QEII Fellowship. Data used here is from the Space Environment Center of
the US National Oceanic and Atmospheric Administration (NOAA). This paper
has benefitted from the comments of two anonymous reviewers.
\end{acknowledgments}

%
%
%
%
%
%
%
%


\vspace{1cm}

%
%
%


%
%
%
\begin{table}
\caption{Verification statistics for the prediction of M-X and of X events 
for 1976-2003}
\begin{flushleft}
\begin{tabular}{lccc}
\tableline
& M-X & & X \\
\cline{2-2} \cline{4-4} \\

Total days & 10226 & & 10226 \\
Event days & 2600 & & 365 \\
$\langle f\rangle$ & 0.282 & & 0.040 \\ 
$\langle x\rangle$ & 0.254 & & 0.036 \\
Median $f$ & 0.218 & & 0.017 \\
$\sigma_f$ & 0.251 & & 0.058 \\
$\sigma_x$ & 0.435 & & 0.186 \\
$\langle f|x=1\rangle$ & 0.508 & & 0.120 \\ 
$\langle f|x=0\rangle$ & 0.204 & & 0.038 \\
Median $f|x=1$ & 0.565 & & 0.105 \\
Median $f|x=0$ & 0.116 & & 0.016 \\
Std.\ dev.\ $f|x=1$ & 0.216 & & 0.084 \\
Std.\ dev.\ $f|x=0$ & 0.212 & & 0.055 \\ 
${\rm MAE}(f,x)$ & 0.277 & & 0.068 \\
${\rm MSE}(f,x)$ & 0.138 & & 0.032 \\
Linear association & 0.528 & & 0.262 \\
${\rm SS}(f,x)$ & 0.272 && 0.066 \\
\tableline
\end{tabular}
\end{flushleft}
\end{table}

\begin{table}
\caption{Dependence of prediction of M-X and of X events on free parameters}
\begin{flushleft}
\begin{tabular}{cccccc}
\tableline
\verb+PRIOR+ & $S_1$ & $T$ & Rate & $\langle \epsilon_{\rm M\!X}\rangle$ 
  & $\langle \epsilon_{\rm X}\rangle$ \\
& (${\rm W}\,{\rm m}^{-2}$) & (years) & prior? & & \\
\tableline
2 & $4\times 10^{-6}$ & 1 & yes & 0.282 & 0.040 \\ 
4 & $4\times 10^{-6}$ & 1 & yes & 0.280 & 0.040 \\ 
2 & $2\times 10^{-6}$ & 1 & yes & 0.286 & 0.052 \\ 
2 & $4\times 10^{-6}$ & 2 & yes & 0.285 & 0.041 \\ 
2 & $4\times 10^{-6}$ & 1 & no & 0.289 & 0.069 \\ 
\tableline
& & & Observed: & 0.254 & 0.036 \\
\end{tabular}
\end{flushleft}
\end{table}

\begin{table}
\caption{Comparison of predictions of present method with NOAA 
predictions, 1987-2003}
\begin{flushleft}
\begin{tabular}{lccccc}
\hline
&\multicolumn{2}{c}{Present} & & \multicolumn{2}{c}{NOAA} \cr
&\multicolumn{2}{c}{method} & & \multicolumn{2}{c}{method} \cr
\cline{2-3} \cline{5-6} \\
& M-X & X & & M-X & X \\
\hline 
$\langle f\rangle$ & 0.294 & 0.040 & & 0.298 & 0.064 \\
$\langle x\rangle $ & 0.262 & 0.035 & & 0.262 & 0.035 \\
$\langle f|x=1\rangle $ & 0.510 & 0.122 & & 0.551 & 0.244 \\
$\langle f|x=0\rangle $ & 0.217 & 0.037 & & 0.208 & 0.057 \\
${\rm MAE}(f,x)$ & 0.289 & 0.066 & & 0.271 & 0.081 \\ 
${\rm MSE}(f,x)$ & 0.143 & 0.031 & & 0.139 & 0.032 \\
${\rm SS}(f,x)$ & 0.258 & 0.078 & & 0.262 & -0.006 \\
\hline
\end{tabular}
\end{flushleft}
\end{table}

%
%
%

%
%
%
%
%
%
%
%
%
%

%
%

\end{article}

\end{document}